\documentclass[USenglish,twocolumn]{article}

\usepackage[utf8]{inputenc}
\usepackage[big]{dgruyter_author}
 
\usepackage{graphicx}
\usepackage{bm}
\usepackage{color}
\usepackage{subfigure}

\begin{document}

  \articletype{Research Article{\hfill}Open Access}

  \author*[1]{Francesco V. Pepe}
	
	\author[2]{Giuliano Scarcelli}

  \author[3]{Augusto Garuccio}
	
	\author[4]{Milena D'Angelo}

  \affil[1]{Museo Storico della Fisica e Centro Studi e Ricerche ``Enrico Fermi'', I-00184 Roma, Italy, and INFN, Sezione di Bari, I-70126 Bari, Italy, E-mail: francesco.pepe@ba.infn.it}

	\affil[2]{Fischell Department of Bioengineering, University of Maryland, College Park MD 20742 USA, E-mail:  
scarc@umd.edu}
	
	\affil[3]{Dipartimento Interateneo di Fisica, Universit\`a degli studi di Bari, and INFN, Sezione di Bari, I-70126 Bari, Italy. E-mail: augusto.garuccio@uniba.it}

\affil[4]{Dipartimento Interateneo di Fisica, Universit\`a degli studi di Bari, and INFN, Sezione di Bari, I-70126 Bari, Italy. E-mail: milena.dangelo@uniba.it}

  \title{\huge Plenoptic imaging with second-order correlations of light}

  \runningtitle{Plenoptic imaging with second-order correlations}


  \begin{abstract}
{Plenoptic imaging is a promising optical modality that simultaneously captures the location and the propagation direction of light in order to enable tridimensional imaging in a single shot. We demonstrate that it is possible to implement plenoptic imaging through second-order correlations of chaotic light, thus enabling to overcome the typical limitations of classical plenoptic devices.}

\end{abstract}

  \journalname{Quantum Measurements and Quantum Metrology}

  \startpage{1}

\maketitle

\section{Introduction}

Plenoptic imaging captures information on the three-dimensional lightfield of a given scene by recording, in a single shot, both the location and the propagation direction of the incoming light \cite{adelson}. The main advantage of recording the propagation direction is the possibility to computationally retrace, in post processing, the geometrical light path, in order to refocus  different planes within the given scene and to extend the depth of field. The key feature of a plenoptic imaging device is a \textit{microlens array} inserted in the native image plane, between the imaging lens and the sensor. The microlenses act as imaging pixels to collect spatial information of the scene. Moreover, each one of them reproduces an image of the main lens on the sensor array, thus providing the angular information associated with each imaging pixel \cite{ng}. Despite being very useful for extending the depth of field, such a structure entails a strong trade-off between spatial and angular resolution.

Digital cameras enhanced by refocusing capabilities make use of plenoptic imaging \cite{website}, thus simplifing both auto-focus and low-light shooting \cite{ng}. A plethora of innovative applications, from 3D-imaging \cite{3dimaging}, to stereoscopy \cite{adelson,muenzel,levoy}, and microscopy \cite{microscopy1,microscopy2,microscopy3,microscopy4} are also being developed. Despite being very useful for extending the depth of field, the structure of plenoptic imaging devices entails a strong trade-off between spatial and angular resolution, in the form of an inverse proportionality. Attempts to decouple resolution and depth of field, based on signal processing, have been proposed in literature \cite{microscopy2,microscopy4,waller,spatioangular,imageformation,superres}. 

We have recently proposed \cite{CPI_PRL} to improve the performances of plenoptic imaging by merging it with ghost imaging \cite{pittman,gatti2,laserphys,valencia,scarcelliPRL,ferri}. The phenomenon of ghost imaging is  typical of correlated light sources, such as entangled photons and chaotic light. Its peculiarity is to enable retrieving the image of a remote object by measuring coincidences/correlations between two separate detectors. 
Such traditional ghost imaging scheme can be modified to enable plenoptic imaging \cite{CPI_PRL}: Using the intrinsic momentum and position correlation of either entangled or chaotic light sources, one can perform imaging in one arm and simultaneously obtain the direction information in the other arm. In particular, we have shown that the second-order correlation function of chaotic light features a nontrivial part that possesses plenoptic properties, thus yielding the possibility to refocus and extend the depth of field of the acquired image. Correlation plenoptic imaging has the advantage of inducing a weaker coupling between spatial and angular resolution, enabling one to reach larger depths of field at fixed resolution.

In this paper, we present a detailed theoretical analysis of correlation plenoptic imaging with chaotic light. In particular, in Sect.\ref{basic} we introduce the proposed setup and its theoritical description, emphasizing the differences with respect to standard chaotic ghost imaging. In Sect.\ref{CPI}, we discuss the plenoptic properties of the second-order correlation function, and show the refocusing enabled by correlation plenoptic imaging. In Sect.\ref{camera}, we discuss the improvements enabled by correlation plenoptic imaging, with respect to standard imaging, in view of the implementation of a correlation plenoptic camera. Conclusions and future perspectives are discussed in Sect.\ref{concl}.

\section{Plenoptic imaging with two-photon correlations}\label{basic}

\begin{figure}
\centering
\includegraphics[width=0.49\textwidth]{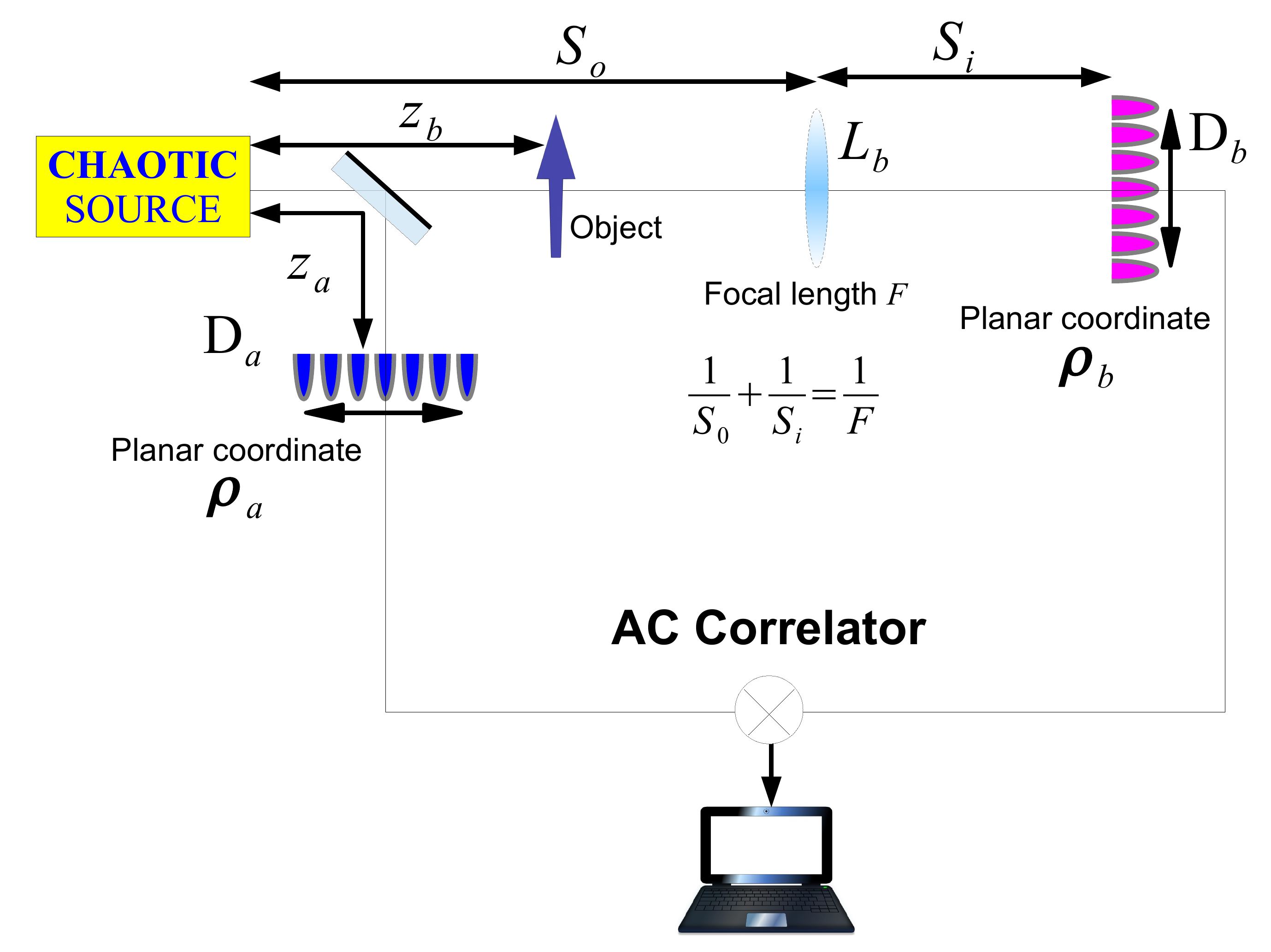}
\caption{Proposed setup for achieving plenoptic imaging by second-order correlation measurements. The lens $L_b$ in the transmitted arm makes an image of the chaotic source on the detector array $\mathrm{D}_b$. A ghost image of the object can be retrieved on $\mathrm{D}_a$, placed in the reflected arm, by measuring second-order correlations. The signals from the two detectors are correlated in AC to suppress the trivial part of the second-order correlation function (\ref{glauber}).}\label{fig:setup}
\end{figure}

\begin{figure}
\includegraphics[width=0.39\textwidth]{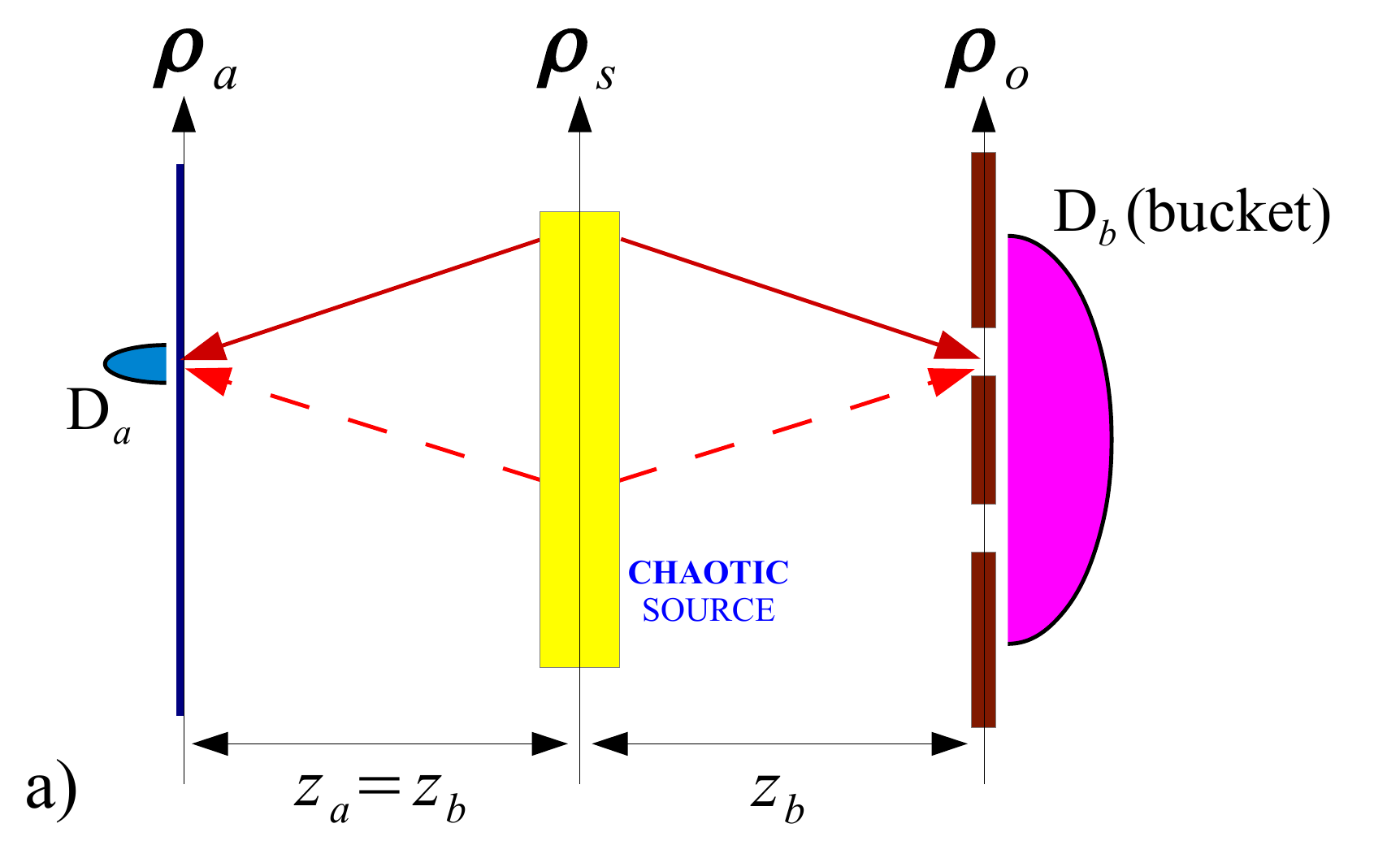} 
\includegraphics[width=0.49\textwidth]{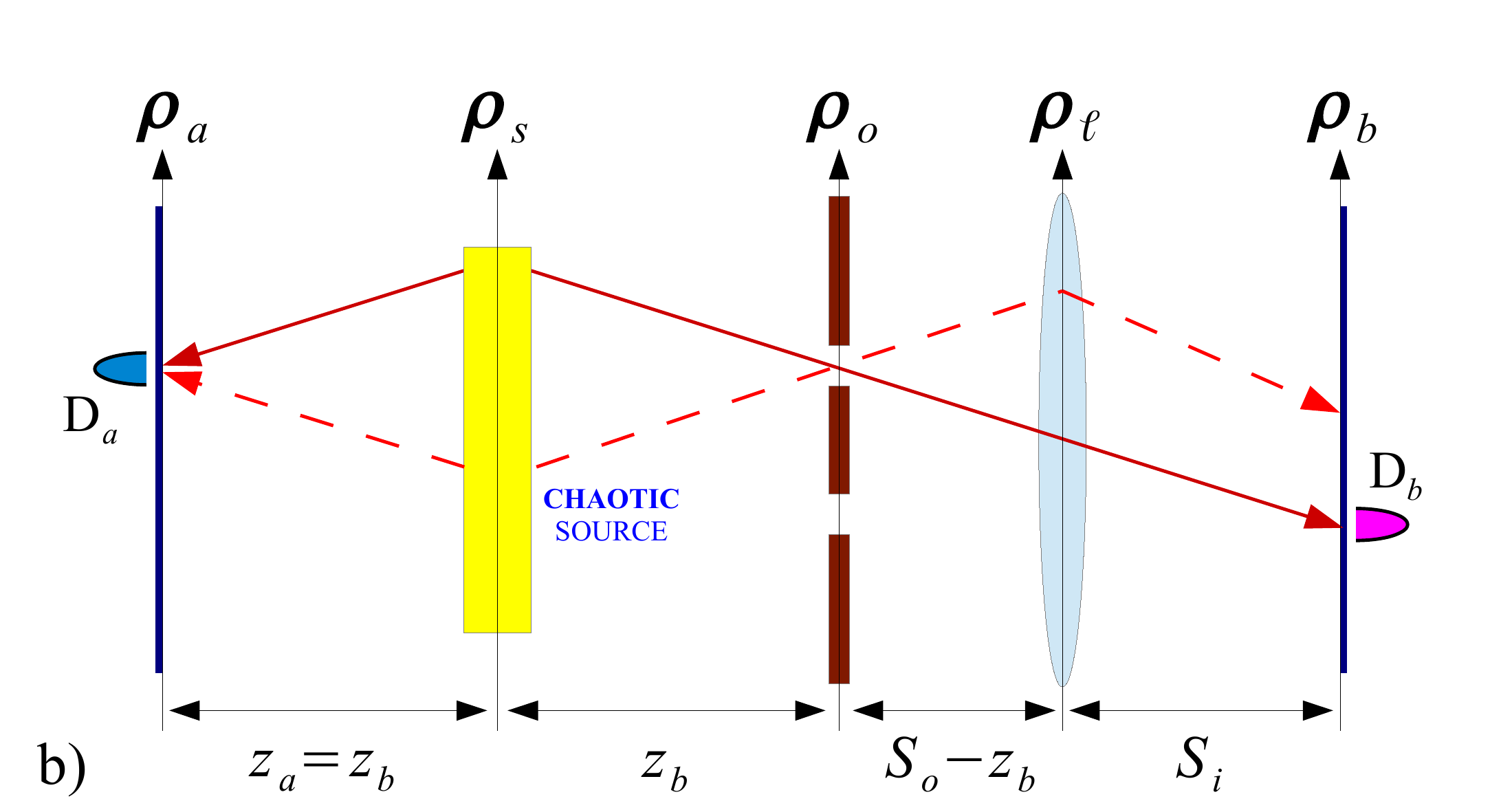}
\caption{Comparison between the unfolded setup in standard chaotic ghost imaging (a) and in chaotic correlation plenoptic imaging (b). Both pictures represent focused setups with $z_a=z_b$.  In the case of ghost imaging (a), the sensor $\mathrm{D}_b$ is a bucket detector collecting light trasmitted by the entire object, with no spatial or directional resolution. In correlation plenoptic imaging, the correlation between the signals from the point-detectors $\mathrm{D}_a$ and $\mathrm{D}_b$ enables to reconstruct the trasmission function of the object and to detect, at the same time, the direction of the light emitted by the source. }\label{fig:klyshko}
\end{figure}

We shall study the properties of second-order correlation in the setup shown in Fig.~\ref{fig:setup}. Light from a chaotic source is split by a symmetric non-polarizing beam splitter (BS) in a reflected arm ($a$), with the detector array $\mathrm{D}_a$ at an optical distance $z_a$ from the source, and a transmitted arm ($b$), in which light impinges on the object to be imaged. The object is placed at a distance $z_b$ from the source. In arm $b$, a thin lens $L_b$ of focal length $F$ focuses on the detector array $\mathrm{D}_b$ light coming from the source and transmitted by the object. Spatio-temporal correlation is then measured between the light intensities detected by each pixel of the two detector arrays. It is known that correlation measurements between the entire sensor $\mathrm{D}_b$ and each pixel of the sensor $\mathrm{D}_a$ enables retrieving the \textit{ghost image} of the object \cite{gatti2,laserphys,valencia,scarcelliPRL,ferri}, which will be perfectly focused if $z_a=z_b$; in fact, the chaotic source acts as a focusing element, as emphasized by the unfolded setup reported in Fig.~\ref{fig:klyshko}a) cite{laserphys,valencia}. For standard ghost imaging, the high-resolution detector array $\mathrm{D}_b$ is actually redundant, since a \textit{bucket detector} behind the object suffice to retrieve the ghost image, as shown in Fig.~\ref{fig:klyshko}a). However, for correlation plenoptic imaging, the detector array $\mathrm{D}_b$ is crucial: It enables to capture information on the propagation direction of light passing through the object. In fact, we will show that spatial and angular information are simultaneously encoded in the second-order correlation function, thus enabling to refocus the image acquired at $z_a\neq z_b$. 

The second-order correlation measurement between each pixel of the sensors $\mathrm{D}_a$ and $\mathrm{D}_b$ is described by the Glauber correlation function \cite{scully}
\begin{eqnarray}\label{glauber}
G^{(2)}(\bm{\rho}_a,\bm{\rho}_b;t_a,t_b)\! & \!=\! & \!\left\langle E^{(-)}_a
(\bm{\rho}_a,t_a) E^{(-)}_b (\bm{\rho}_b,t_b) \right. \nonumber
\\ & & \left. E^{(+)}_b (\bm{\rho}_b,t_b) E^{(+)}_a
(\bm{\rho}_a,t_a) \right\rangle,
\end{eqnarray}
where $\bm{\rho}_{i}$ indicates the planar coordinate on the sensor $\mathrm{D}_{i}$ (with $i=a,b$), $t_{i}$ is the corresponding detection time, and $E_{i}^{(\pm)}$ are the positive- and negative-frequency components of the electric field $E^{(+)}=(E^{(-)})^{\dagger}$ at each detector, for which a scalar approximation is assumed. The expectation value $\langle O \rangle= \mathrm{Tr}(\varrho O)$ in Eq.(\ref{glauber}) is evaluated by considering the quantum state $\varrho$ of the source. The fields at the detectors can be expressed in terms of the field on the output plane of the source by using the optical transfer functions $g_i(\bm{k})$ in the paraxial approximation \cite{goodman}, which is
\begin{equation}\label{field}
E^{(+)}_i (\bm{\rho}_i,t_i) = C \int d\omega \int d\bm{\kappa}\, a_{\bm{k}} e^{-i\omega t_i} g_i(\bm{\rho}_i,\bm{k}),
\end{equation}
where $C$ is a normalization constant, $\omega$ is the frequency, $\bm{\kappa}$ is the transverse momentum, $a_{\bm{k}}$ is the canonical field operator, associated with the mode $\bm{k}$, which satisfies the commutation relation: $[a_{\bm{k}},a_{\bm{k}'}^{\dagger}]=\delta_{\bm{k},\bm{k}'}$. The three-dimensional wave vector $\bm{k}=(\bm{\kappa},k_z)$ is such that $k_z\simeq \omega/c$. After inserting in Eq.(\ref{glauber}) the fields defined by Eq.(\ref{field}), one finds that, if the source is stationary and quasi-monochromatic, with frequency $\omega_0$, the four-point expectation value $\langle a_{\bm{k}_1}^{\dagger} a_{\bm{k}_2}^{\dagger} a_{\bm{k}_3} a_{\bm{k}_4} \rangle$ is nonvanishing only if the moduli of all momenta are close to $\omega_0/c$. Under these assumptions, the correlation function depends only on $\tau=t_a-t_b$, and the time dependent part factorizes with respect to the spatial part. Moreover, if the source is chaotic, the four-point expectation value is the sum of a ``direct'' and an ``exchange'' term
\begin{equation}
\langle a_{\bm{k}_1}^{\dagger} a_{\bm{k}_2}^{\dagger} a_{\bm{k}_3} a_{\bm{k}_4} \rangle \propto \delta (\bm{k}_1-\bm{k}_4) \delta (\bm{k}_2-\bm{k}_3) + \delta (\bm{k}_1-\bm{k}_3) \delta (\bm{k}_2-\bm{k}_4),
\end{equation}
which are related to the bosonic symmetrization of two-photon states. Therefore, upon neglecting both the time dependence and irrelevant normalization constants, the second-order correlation function defined in Eq.(\ref{glauber}) reads, for a stationary, quasi-monochromatic and chaotic source:
\begin{equation}\label{G2thermal}
G^{(2)}(\bm{\rho}_a,\bm{\rho}_b) = I_a (\bm{\rho}_a) I_b (\bm{\rho}_b) + \Gamma(\bm{\rho}_a,\bm{\rho}_b),
\end{equation} 
where the first term is the mere product of intensities: 
\begin{equation}\label{intensity}
I_i (\bm{\rho}_i) = \int d\bm{\kappa} |g_i(\bm{\rho}_i,\bm{\kappa})|^2
\end{equation}
at the pixel located in $\bm{\rho}_i$  of the sensor $\mathrm{D}_i$; the frequency dependence has been dropped in the $g_i$'s. The second term
\begin{equation}\label{Gamma}
\Gamma(\bm{\rho}_a,\bm{\rho}_b) = \left| \int d\bm{\kappa} g_a(\bm{\rho}_a,\bm{\kappa})^* g_b(\bm{\rho}_b,\bm{\kappa}) \right|^2
\end{equation}
represents the nontrivial part of the second-order correlation. As we shall soon demonstrate, the interesting part $\Gamma(\bm{\rho}_a,\bm{\rho}_b)$ of the second-order correlation function encodes plenoptic imaging properties.

In order to unveil such imaging properties, we first need to compute the transfer functions $g_a$ and $g_b$. To this end, we use the paraxial Gaussian propagator \cite{goodman}
\begin{equation}
\mathcal{G}(\bm{\rho},z;\omega) = G(\bm{\rho})_{\left[\frac{\omega}{cz}\right]} h(\omega,z),
\end{equation}
with
\begin{equation}
G(\bm{\rho})_{[\beta]} = \exp \left( \frac{i}{2} \beta \bm{\rho}^2 \right), \quad h(\omega,z) = -i\frac{\omega}{2\pi c z} e^{i \frac{\omega}{c} z},
\end{equation}
and treat  
the source as an emitter of incoherent paraxial waves. In the reflected arm $a$, light propagates in free space from the source to the detector $\mathrm{D}_a$. Hence, the corresponding transfer function is
\begin{align}\label{propa}
& g_a(\bm{\rho}_a,\bm{\kappa}) = h(\omega_0,z_a) \!\int \!d\bm{\rho}_s f(\bm{\rho}_s) e^{i\bm{\kappa}\cdot\bm{\rho}_s} G(\bm{\rho}_a-\bm{\rho}_s)_{\left[\frac{\omega_0}{cz_a}\right]} \nonumber
\\ & = \mathcal{C}_a(\bm{\rho}_a,z_a) \int \!d\bm{\rho}_s f(\bm{\rho}_s) e^{i\left(\bm{\kappa} - \frac{\omega_0}{cz_a}\bm{\rho}_a \right)\cdot\bm{\rho}_s} G(\bm{\rho}_s)_{\left[\frac{\omega_0}{cz_a}\right]},
\end{align}
where $f(\bm{\rho}_s)$ is the source profile, and
\begin{equation}
\mathcal{C}_a(\bm{\rho}_a,z_a) := h(\omega_0,z_a) G(\bm{\rho}_a)_{\left[\frac{\omega_0}{cz_a}\right]}.
\end{equation}
Computation of the field propagator $g_b$ is slightly more involved, since it requires additional integration on the object and the lens planes, namely:
\begin{align}
& g_b(\bm{\rho}_b,\bm{\kappa}) = h(\omega_0,z_b) h(\omega_0,S_o-z_b) h(\omega_0,S_i) \nonumber \\ & \times\!\int\!d\bm{\rho}_s f(\bm{\rho}_s) e^{i\bm{\kappa}\cdot\bm{\rho}_s} \!\int\!d\bm{\rho}_o G(\bm{\rho}_o-\bm{\rho}_s)_{\left[\frac{\omega_0}{cz_b}\right]} A(\bm{\rho}_o) \nonumber 
\\ & \times\!\int\!d\bm{\rho}_{\ell} G(\bm{\rho}_{\ell}-\bm{\rho}_o)_{\left[\frac{\omega_0}{c(S_o-z_b)}\right]} L(\bm{\rho}_{\ell}) G(\bm{\rho}_b-\bm{\rho}_{\ell})_{\left[\frac{\omega_0}{c S_i}\right]} ,
\end{align}
where $A(\bm{\rho}_o)$ and $L(\bm{\rho}_{\ell})$ are the transmission functions of the object and the lens, respectively. Henceforth, we shall assume that the lens $L_b$ is diffraction-limited, and approximate its transmission function with the Gaussian phase shift $G(\bm{\rho}_{\ell})_{[-\omega_0/cF]}$, where $F$ is the focal length. Assuming that the distance from the source to the lens ($S_o$) and from the lens to the detector ($S_i$) are conjugate (i.e., $1/S_i+1/S_o=1/F$), the propagator in arm $b$ reduces to
\begin{align}\label{propb}
& g_b(\bm{\rho}_b,\bm{\kappa}) = \mathcal{C}_b(\bm{\rho}_b,z_b) \!\int\!d\bm{\rho}_s\!\int\!d\bm{\rho}_o f(\bm{\rho}_s) A (\bm{\rho}_o) \nonumber
\\ & \times G(\bm{\rho}_s)_{\left[\frac{\omega_0}{cz_b}\right]} e^{i\bm{\kappa}\cdot\bm{\rho}_s - \frac{i\omega_0}{cz_b} \bm{\rho}_o\cdot \left( \bm{\rho}_s + \frac{\bm{\rho}_b}{M} \right)} ,
\end{align} 
where $M=S_i/S_o$ is the lens magnification and 
\begin{align}
\mathcal{C}_b(\bm{\rho}_b,z_b) := & h(\omega_0,z_b) h(\omega_0,S_i) \frac{S_o}{z_b} e^{i\frac{\omega_0}{c}(S_o-z_b)} \nonumber \\ & \times G(\bm{\rho}_a)_{\left[\frac{\omega_0}{cS_i} \left( 1- \frac{S_o-z_b}{Mz_b}\right)\right]}.
\end{align}

Given the propagators of Eq.s(\ref{propa})-(\ref{propb}), it is now straightforward to compute the correlation function of Eq.(\ref{G2thermal}). 
In particular, the intensities at the detectors [Eq.(\ref{intensity})] are given by:
\begin{equation}
I_a (\bm{\rho}_a) = \mathcal{K}_a(z_a) \!\int \!d\bm{\rho}_s F(\bm{\rho}_s) ,
\end{equation}
where $F=|f|^2$ the intensity profile of the source and $\mathcal{K}_a = |2\pi\mathcal{C}_a|^2$, and
\begin{equation}
I_b (\bm{\rho}_b) = \mathcal{K}_b(z_b) \!\int \!d\bm{\rho}_s F(\bm{\rho}_s) \left| \tilde{A} \left[ \frac{\omega_0}{cz_b} \left( \bm{\rho}_s + \frac{\bm{\rho}_b}{M} \right) \right] \right|^2
\end{equation}
with $\tilde{A}(\bm{\kappa}) = \int d\bm{\rho}_o A(\bm{\rho}_o) e^{-i\bm{\kappa}\cdot\bm{\rho}_o}$, and $\mathcal{K}_b = |2\pi\mathcal{C}_b|^2$. Therefore,  neither intensity profiles at the sensors, $I_a$ and $I_b$, enable to image the object; in particular, $I_a$ is flat, while the intensity at $\mathrm{D}_b$ is modulated by the squared Fourier transform of the object trasmission function ($\tilde{A}$). In fact, $I_b$ reduces to the squared Fourier transform of the object trasmission funtion, for a point-like source, and to a constant, for an infinitely extended source. 
The crossed part of the second-order correlation function [Eq.(\ref{Gamma})], on which the following analysis will be focused, reads:
\begin{align}\label{Gammazazb}
& \Gamma_{(z_a,z_b)} (\bm{\rho}_a,\bm{\rho}_b) = \mathcal{K}_a(z_a) \mathcal{K}_b(z_b) \!\Biggl|\!\int\!d\bm{\rho}_o\!\int\!d\bm{\rho}_s A(\bm{\rho}_0) \nonumber \\
& \times F(\bm{\rho}_s) G(\bm{\rho}_s)_{\left[\frac{\omega_0}{c}\!\left(\frac{1}{z_b}- \frac{1}{z_a} \right)\!\right]} e^{-\frac{i\omega_0}{c z_b} \!\left[\left( \bm{\rho}_o - \frac{z_b}{z_a} \bm{\rho}_a \right)\! \cdot \bm{\rho}_s + \bm{\rho}_o \cdot \frac{\bm{\rho}_b}{M} \right]} \Biggr|^2 ;
\end{align}
the notation has been chosen to highlight its dependence on the distances $z_{a}$ and $z_{b}$. The imaging properties associated with the result of Eq.(\ref{Gammazazb}) will be discussed in the next section. For the moment, let us observe that integration over the whole sensor  $\mathrm{D}_b$, at fixed $\bm{\rho}_a$, yields an incoherent image of the object, whose point-spread function is determined by the Fourier transform of $F(\bm{\rho}_s)G(\bm{\rho}_s)$. Thus, the minimal point-spread occurs when $z_b=z_a$, and coincides with the typical chaotic ghost imaging. The focused ghost image reads
\begin{align}\label{image}
& \Sigma_{z_a} (\bm{\rho}_a) = \!\int\! d\bm{\rho}_b \Gamma_{(z_a,z_a)} (\bm{\rho}_a,\bm{\rho}_b) \nonumber \\
& \propto \!\int \!d\bm{\rho}_o |A(\bm{\rho}_o)|^2 \left| \tilde{F} \left[ \frac{\omega_0}{cz_a} \left( \bm{\rho}_o - \bm{\rho}_a \right) \right] \right|^2 ,
\end{align}
which sets a quasi one-to-one correspondence between points of the object ($\bm{\rho}_o$) and pixels of the sensor $\mathrm{D}_a$ ($\bm{\rho}_a$), with an uncertainty $\Delta\rho_a \simeq 2\pi c z_a/(\omega_0 D_s) =: \lambda_0 z_a/D_s$ defined by the effective diameter $D_s$ of the source. On the other hand, due to the first-order image of the source on sensor $\mathrm{D}_b$, Eq.~(\ref{Gammazazb}) also entails a correspondence between points of the source plane ($\bm{\rho}_s$) and pixels of the sensor $\mathrm{D}_b$ ($\bm{\rho}_b=-M\bm{\rho}_s$), whose uncertainty $\Delta\rho_b = M \lambda_0 z_b/d$ is determined by the typical size $d$ of the smallest detail of the object, which acts as a pupil for the lens $L_b$. The Klyshko-like picture \cite{klyshko} representing the double focusing effect encoded in $\Gamma_{(z_a,z_a)}$, as compared with standard chaotic ghost imaging, is represented in Figure~\ref{fig:klyshko}. Notice that, based on the structure of the correlation term of Eq.(\ref{Gamma}), the source acts as a phase conjugate mirror, hence the correlated modes in the two arms of the setup are characterized by identical transverse momenta.

\section{Refocusing and changing the point of view}\label{CPI}

To unveil the plenoptic properties encoded in the correlation term $\Gamma_{(z_a,z_b)}$ of Eq.(\ref{Gammazazb}), it is worth resorting to the geometrical optics limit $\omega_0\to\infty$. Let us first observe that the double integral in Eq.(\ref{Gammazazb}) has the form
\begin{equation}
\int\!d\bm{\rho}_o d\bm{\rho}_s A(\bm{\rho}_0) F(\bm{\rho}_s) e^{i\frac{\omega_0}{c} \varphi(\bm{\rho}_o,\bm{\rho}_s;\bm{\rho}_a,\bm{\rho}_b)} ,
\end{equation}
in which the aperture function of the object and the intensity profile of the source are linked by the phase
\begin{equation}\label{phi}
\varphi(\bm{\rho}_o,\bm{\rho}_s;\bm{\rho}_a,\bm{\rho}_b) = \frac{\bm{\rho}_s^2}{2}\!\left(\frac{1}{z_b}-\frac{1}{z_a}\right) - \frac{\bm{\rho}_o}{z_b}\cdot \!\left(\bm{\rho}_s + \frac{\bm{\rho}_b}{M}\right) + \frac{\bm{\rho}_s\cdot\bm{\rho}_a}{z_a}.
\end{equation}
In the short-wavelength limit, the integral is dominated by the stationary points of the phase with respect to the integration variables. In particular, by imposing the stationarity of the function defined in Eq.(\ref{phi}) with respect to $\bm{\rho}_o$, we get
\begin{equation}\label{cond1}
\bm{\rho}_s + \frac{\bm{\rho}_b}{M} = 0,
\end{equation}
which gives the geometrical correspondence between points of the source and points of the sensor $D_b$, with magnification $M$ and inversion of the image. The stationarity condition with respect to $\bm{\rho}_s$ is much less trivial, since it involves the source, the object and the detector $D_a$, namely: 
\begin{equation}\label{cond2}
\bm{\rho}_s \left(\frac{1}{z_b}-\frac{1}{z_a}\right) - \frac{\bm{\rho}_o}{z_b} + \frac{\bm{\rho}_a}{z_a} =0.
\end{equation}
After substituting Eq.(\ref{cond1}) into Eq.(\ref{cond2}), we find that the point of the object  yielding the dominant contribution to the integral, at the given detection positions/pixels, reads
\begin{equation}
\bm{\rho}_o = \frac{z_b}{z_a} \bm{\rho}_a - \frac{\bm{\rho}_b}{M} \left( 1-\frac{z_b}{z_a} \right).
\end{equation}
Thus, in the geometrical optics limit, the nontrivial part of the second-order correlation function asymptotically behaves like
\begin{equation}\label{scaling0}
\Gamma_{(z_a,z_b)} (\bm{\rho}_a,\bm{\rho}_b) \sim F\!\left( - \frac{\bm{\rho}_b}{M} \right)^2 \!\left| A\!\left[ \frac{z_b}{z_a} \bm{\rho}_a - \frac{\bm{\rho}_b}{M} \left( 1-\frac{z_b}{z_a} \right) \right] \right|^2 ,
\end{equation}
which is, it contains information on boh the object and the source. If $z_b\neq z_a$, the integration over $\rho_b$ [similar to the one in Eq.(\ref{image})] would cancel the information on the aperture function of the object, thus making it impossible to retrieve the ghost image. This indicates the crucial role played by the high-resolution detector $D_b$, as opposed to the bucket detector of standard ghost imaging, for plenoptic imaging purposes. In fact, the out-of-focus image obtained by the correlation measurement can be correctly reconstructed (i.e., \textit{refocused}) by employing the following scaling property:
\begin{equation}\label{scaling}
\Gamma_{(z_a,z_b)} \left[\frac{z_a}{z_b} \bm{\rho}_a - \frac{\bm{\rho}_b}{M} \left( 1-\frac{z_a}{z_b} \right),\bm{\rho}_b\right] \sim F\!\left( - \frac{\bm{\rho}_b}{M} \right)^2 \!\left| A(\bm{\rho}_a) \right|^2 .
\end{equation}
This scaling property is formally identical to the one employed in plenoptic imaging \cite{ng}. In fact, as in standard plenoptic imagig, refocusing is enabled by the information on the propagation direction of the light which contributed to the image formation. In our scheme, correlation measurement enables identifying the source point from which light has been emitted, and thus to reconstruct the signal ``trajectory'' from the source to the detector, through the object [see Figure~\ref{fig:klyshko}b)]. The refocused image of Eq.(\ref{scaling}), which would in principle suffice for high-depth-of-field imaging, is typically affected by a low signal-to-noise ratio. To overcome this problem, it is convenient to integrate the refocused image over the whole sensor $\mathrm{D}_b$, thus obtaining:
\begin{align}
\Sigma^{\mathrm{ref}}_{(z_a,z_b)} (\bm{\rho}_a) & = \!\int\!d\bm{\rho}_b \Gamma_{(z_a,z_b)} \left[\frac{z_a}{z_b} \bm{\rho}_a - \frac{\bm{\rho}_b}{M} \left( 1-\frac{z_a}{z_b} \right),\bm{\rho}_b\right] \nonumber \\
& \simeq \frac{\mathcal{K}_b(z_b)}{\mathcal{K}_b(z_a)} \Sigma_{z_a} (\bm{\rho}_a),
\end{align}
which is, up to an intensity-rescaling factor, the incoherent image of Eq.(\ref{image}):

The result in Eq.~(\ref{scaling0}) is related to another relevant imaging property, of practical interest for 3D imaging. In fact, in an out-of-focus image, different pixel coordiantes $\bm{\rho}_b$ correspond to different point of views on the object. Each pixel on the sensor $\mathrm{D}_b$ represents a different point of view from which the image of the object (focused in the image plane $z_a=z_b$) is projected onto the sensor $\mathrm{D}_a$ at $z_a\neq z_b$.

In reality, due to diffraction, the correlation image tends to increasingly spread as the object is more and more out of focus. To better see this point, let us consider a source with a Gaussian intensity profile
\begin{equation}
F(\bm{\rho}_s) = \frac{1}{2\pi\sigma^2} \exp \!\left( - \frac{\bm{\rho}_s^2}{2 \sigma^2} \right);
\end{equation}
in this case, the \textit{coherent} point-spread function appearing in $\Gamma_{(z_a,z_b)}$ [see Eq.~(\ref{Gammazazb})] reduces to
\begin{align}\label{PSFc}
& \!\int\!d\bm{\rho}_s F(\bm{\rho}_s) G(\bm{\rho}_s)_{\left[\frac{\omega_0}{c z_b}\!\left(1- \alpha \right)\!\right]} e^{-\frac{i\omega_0}{c z_b} \!\left( \bm{\rho}_o - \alpha \bm{\rho}_a \right)\! \cdot  \bm{\rho}_s } \nonumber \\
& \propto \exp\!\left( -\frac{1}{2} \!\left( \frac{\omega_0\sigma}{c z_b} \right)^2\! \frac{|\bm{\rho}_0-\alpha\bm{\rho}_a|^2}{1 - \frac{i\omega_0\sigma^2}{cz_b}(1-\alpha)} \right) ,
\end{align}
with $\alpha=z_b/z_a$. Now, the point-spread function of the incoherent image is given by the square modulus of the result in Eq. (\ref{PSFc}), namely,
\begin{equation}\label{PSFi}
\exp\!\left( - \!\left( \frac{\omega_0\sigma}{c z_b} \right)^2\! \frac{|\bm{\rho}_0-\alpha\bm{\rho}_a|^2}{1 + \left(\frac{\omega_0\sigma^2}{cz_b}(1-\alpha)\right)^2} \right).
\end{equation}
The comparison of these two expressions clarifies the origin of the larger depth of field characterizing the coherent (plenoptic) ghost image, with respect to the incoherent ghost image (which is, the standad ghost image obtained by either using a bucket detector or integrating the choerent image over $\bm{\rho}_b$). In fact, in the geometrical optics limit, the (real) variance of the incoherent point-spread function [Eq. (\ref{PSFi})] approaches the value $\sigma|1-\alpha|$, thus becoming independent of $\omega_0$ . On the contrary, for high frequencies, the variance of the coherent point-spread function [Eq. (\ref{PSFc})] becomes imaginary and eventually vanishes like $\sqrt{cz_b|1-\alpha|/\omega_0}=\sqrt{2\pi\lambda_0 z_b|1-\alpha|}$ as $\omega_0\to\infty$. Thus, provided $\lambda_0 z_b/\sigma^2\ll 1$, the depth of field of the coherent images is much larger than the one obtained with mere ghost imaging.

\section{Towards a correlation plenoptic camera}\label{camera}

In a plenoptic imaging device, the sensor is divided into {\it macropixels} of width $\delta_x$, which define the image resolution. Each macropixel is made of $N_u^\mathrm{(p)}$ (micro)pixels per side, whose width $\delta$ fixes the angular resolution \cite{adelson,ng}. An array of microlenses of diameter $\delta_x$ and focal length $F$ is inserted in front of the sensor for reproducing, within each macropixel, the image of the main camera lens. Hence, each micropixel behind a given microlens collects light from a sector of the main lens, thus encoding information on the direction of light collected by the specific microlens (which corresponds to a specific part of the acquired image). Assuming the sensor to have width $W$, this configuration yields the following inverse proportionality between the number of pixels per side devoted to the spatial ($N_x^\mathrm{(p)}=W/\delta_x$) and to the angular ($N_u^\mathrm{(p)}=\delta_x/\delta$) detection of the lightfield:
\begin{equation}\label{Np}
N_x^{\mathrm{(p)}} N_u^{\mathrm{(p)}} = \frac{W}{\delta} \equiv
N_{\mathrm{tot}},
\end{equation}
where $N_{\mathrm{tot}}$ is the total number of pixels per side on the sensor. Simple geometrical considerations indicate that the maximum achievable depth of field is determined by $N_u^\mathrm{(p)}$. Thus, based on Eq.~(\ref{Np}), the depth of field can be increased only at the expenses of the image resolution $N_x^\mathrm{(p)}$. 

\begin{figure}
\centering
\includegraphics[width=0.35\textwidth]{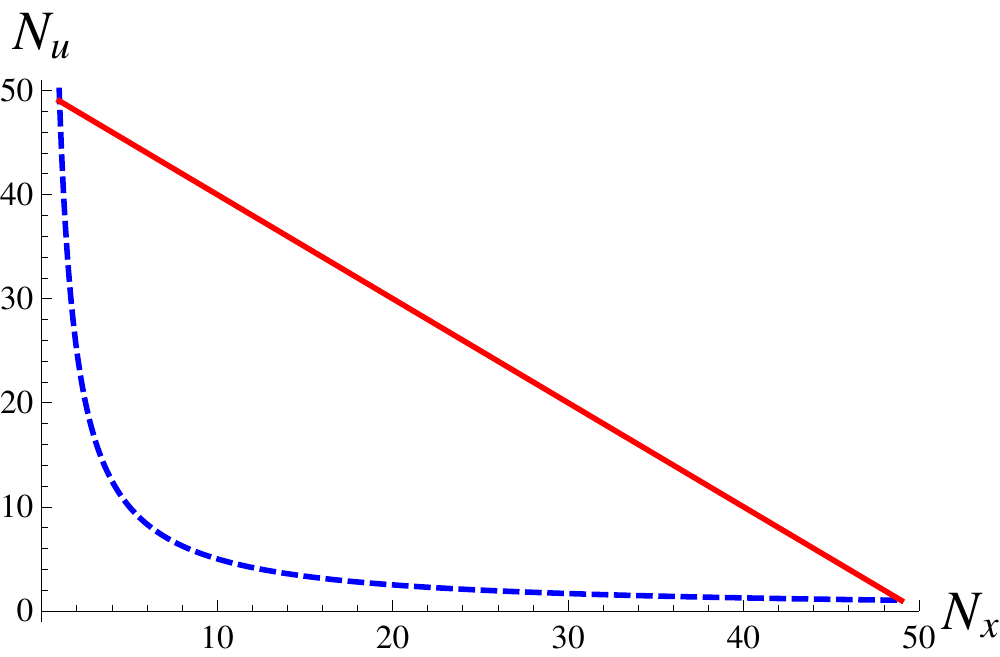}
\caption{The values of $N_x$ and $N_u$ achievable with a standard
plenoptic camera (blue dashed line) and a plenoptic correlation-imaging camera (red solid line), for$N_{\mathrm{tot}}=50$.}\label{fig:NxNu}
\end{figure}

In correlation plenoptic imaging, light is split in two arms, and two spatially decoupled sensors are employed, one for imaging and one for angular measurement. Hence, the two sensors can have the same pixel size $\delta$. Assuming that their widths $W_x^\mathrm{(cp)}$ and $W_u^\mathrm{(cp)}$ satisfy the condition $W_x^\mathrm{(cp)}+W_u^\mathrm{(cp)}=W$, the number of pixels per side dedicated to the spatial and to the directional measurement ($N_i^\mathrm{(cp)}=W_i^\mathrm{(cp)}/\delta$, with $i=x,u$) is constrained by the relation:
\begin{equation}\label{Nc}
N_x^{\mathrm{(cp)}} + N_u^{\mathrm{(cp)}} = N_{\mathrm{tot}}.
\end{equation}
Hence, the depth of field, which depends on $N_u^{\mathrm{(cp)}}$, is characterized by a striking improvement with respect to standard plenoptic imaging, at fixed image resolution of correlation plenoptic imaging. The improvement is larger as $N_{\mathrm{tot}}$ increases (see Figure~\ref{fig:NxNu}).

\section{Conclusions}\label{concl}

We have analyzed the plenoptic properties of correlation imaging, which enable to refocus even largely out-of-focus images. In view of practical applications, it is worth mentioning that the obtained results do not depend on the nature of the object, whether reflective or transmissive. It is also reasonable to expect that such an imaging procedure can be extended to any other sources, either photons or particles \cite{ghost_muons}, that is characterized by correlation in both momentum \textit{and} position \cite{laserphys,gatti2}. For instance, the light source can still be imaged on $\mathrm{D}_b$ when replacing the chaotic source with an entangled photon source, such as SPDC \cite{klyshko}, to obtain the angular information. In this case, a lens is required to achieve ghost imaging of the object \cite{pittman,laserphys}. On the other hand, we do not expect CPI to work with classically correlated beams \cite{bennink} characterized solely by momentum correlation \cite{laserphys}.

\section*{Acknowledgments}

This work has been supported by P.O.N.~RICERCA E COMPETITIVITA' 2007-2013 - Avviso n.~713/Ric.~del 29/10/2010, Titolo II - ``Sviluppo/Potenziamento di DAT e di LPP'' (project n.~PON02-00576-3333585).

\end{document}